\documentclass[%
aps,
pre,
 amsmath,amssymb,
reprint,
]{revtex4-1}
\usepackage{graphicx}
\usepackage{dcolumn}
\usepackage{bm}

\usepackage{amsmath}	
\begin{document}

\preprint{AIP/123-QED}

\title{Universality of classical and quantum SAT-UNSAT transitions of
convex continuous satisfaction problems}

\author{Harukuni Ikeda}
 \email{harukuni.ikeda@gakushuin.ac.jp}
\affiliation{ 
Department of Physics, Gakushuin University, 1-5-1 Mejiro, Toshima-ku, Tokyo 171-8588, Japan}

\date{\today}

\begin{abstract}
Here we investigate the single-layer linearized perceptron near the
SAT-UNSAT transition point as a prototypical model of the convex
continuous satisfaction problems.  The simplicity of the model allows us
to take into account the effects of the quantum fluctuation, which have
not been fully investigated before. We found that the classical and
quantum models have different critical exponents and thus have different
universality classes. We also briefly discuss the effects of the random
field.
\end{abstract}

\maketitle

\newcommand{\diff}[2]{\frac{d#1}{d#2}}
\newcommand{\pdiff}[2]{\frac{\partial #1}{\partial #2}}
\newcommand{\fdiff}[2]{\frac{\delta #1}{\delta #2}}
\newcommand{\bx}{\bm{x}}
\newcommand{\bxi}{\bm{\xi}}
\newcommand{\br}{\bm{r}}
\newcommand{\bu}{\bm{u}}
\newcommand{\tl}{\tilde{\lambda}}
\newcommand{\new}{\nonumber\\}
\newcommand{\abs}[1]{\left|#1\right|}
\newcommand{\tr}{{\rm Tr}}
\newcommand{\HH}{{\mathcal H}}
\newcommand{\II}{{\mathcal I}}
\newcommand{\WW}{{\mathcal W}}
\newcommand{\OO}{{\mathcal O}}
\newcommand{\ave}[1]{\left\langle #1 \right\rangle}
\newcommand{\im}{{\rm Im}}
\newcommand{\re}{{\rm Re}}
\newcommand{\ke}{k_{\rm eff}}
\newcommand{\ipr}{{\rm IPR}}

\section{Introduction} 
The purpose of the constraint satisfaction problem is to find out
solutions satisfying given constraints. There are many such solutions if
the number of constraints is sufficiently smaller than the number of
degrees of freedom. On increasing the number of constraints, the number
of solutions decreases, and eventually, the solution ceases to exist at
a certain point. This is the so-called satisfaction (SAT) unsatisfaction
(UNSAT)
transition~\cite{kirkpatrick1983optimization,gardner1989optimal,franz2017}.
The SAT-UNSAT transition becomes a genuine phase transition in the
thermodynamic limit where both the number of degrees of freedom and the
number of constraints go to infinity~\cite{nishimori2001statistical}.

Recently, the SAT-UNSAT transition of continuous variables has attracted
much attention in connection with the sphere packing
problem~\cite{parisi2010,kurchan2012exact,kurchan2013exact,charbonneau2014exact,charbonneau2014fractal,parisi2020theory,franz2016simplest,franz2017,brito2018universality,franz2019jamming,franz2019critical,franz2019impact,artiaco2021}.
The sphere packing problem is thought as a constraint satisfaction
problem to find a configuration with the constraint that spheres do not
overlap~\cite{krzakala2007,mari2009}. The packing fraction at which such
configurations cease to exist is called the jamming transition point
$\varphi_J$~\cite{ohern2003,krzakala2007,parisi2020theory}.  Several
physical quantities, such as the contact number, shear modulus,
correlation length, and relaxation time, exhibit the critical behavior
near
$\varphi_J$~\cite{ohern2003,ikeda2013dynamic,shimada2018,ikeda2020universal}.

A promising way to study phase transitions is first to consider a
solvable mean-field model~\cite{nishimori2010elements}.  In the case of
the SAT-UNSAT transition of the continuous constraint satisfaction
problem, a prototypical mean-field model is the single-layer
perceptron~\cite{rosenblatt1958perceptron,gardner1989optimal}. For the
classical model of the perceptron, the static critical properties of the
model have already been well investigated by using the replica
method~\cite{franz2016simplest,franz2017,brito2018universality,franz2019jamming,franz2019critical,franz2019impact,artiaco2021}.
In particular, a non-convex version of the model is shown to have the
same critical exponents of those of particle systems near the jamming
transition point~\cite{franz2016simplest,franz2017}. However, due to
the complexity of the model, there are several unsolved problems, even
for the convex case, where the cost function has a unique minimum. For
instance, out-of-equilibrium dynamics of the perceptron has been
actively studied recently because of its relevance to machine learning
and the jamming
transition~\cite{agoritsas2018out,hwang2020,manacorda2022gradient,folena2022introduction}.
However, the current formalism based on the dynamical density functional
theory is numerically highly demanding, which makes it difficult to
estimate the dynamical critical exponent~\cite{folena2022introduction}.
Another theoretically interesting question is how the quantum
fluctuation affects the nature of the SAT-UNSAT transition of the
continuous satisfaction problem. Unfortunately, a quantum version of the
peceptron is difficult to solve analytically and previous studies relied
on the Schehr–Giamarchi–Le Doussal Expansion~\cite{franz2019impact} or
Monte Carlo sampling~\cite{artiaco2021}. In particular, the scaling
behavior of the quantum SAT-UNSAT transition in the UNSAT side has not
been fully investigated
yet~\cite{franz2019impact,artiaco2021}. Considering those difficulties
of the perceptron, it is desirable to first consider a more analytically
tractable model.

In this work, we revisit a simplified problem: a linealized version of
the perceptron~\cite{hertz1989dynamics,hertz1989phase}. The model can be
solved analytically without relying on the replica
method~\cite{hertz1989phase}. Furthermore, its dynamical properties have
been already well
investigated~\cite{hertz1989dynamics,cun1991,watkin1993}. We first
investigate the classical version of the model near SAT-UNSAT transition
point. The model shows the same scaling as that of the original
perceptron of the convex case~\cite{franz2017}. Next, to demonstrate the
usefulness of the model, we consider the quantum version of the
model. We show that the susceptibility against quantum fluctuation
behaves qualitatively differently from thermal fluctuation. Finally, we
briefly discuss the effects of the random field.

This paper is organized as follows. In Sec.~\ref{145208_17Aug22}, we
investigate the classical model.  We characterize the criticality in
terms of the condensation transition as previously done for $p=2$-spin
spherical model~\cite{barbier2022generalised}. In
Sec.~\ref{145228_17Aug22}, we investigate the quantum model.  In
Sec.~\ref{145448_17Aug22}, we investigate the model with a random field.
Finally, In Sec.~\ref{145458_17Aug22}, we summarize the work and discuss
possible future works.

\section{Classical model}
\label{145208_17Aug22} We consider the following continuous constraint
satisfaction problem.  Let $\bx=\{x_1,\dots, x_N\}$ be the state vector
of norm $\bx\cdot\bx=N$. The problem is if there exists $\bx$ such that
\begin{align}
\bx\cdot \bxi^\nu = 0\ {\rm for}\ \nu = 1,\dots, M,\label{100328_16Aug22}
\end{align}
where $\bxi^\nu=\{\xi_1^\nu,\dots, \xi_N^\nu\}$ denotes a $N$
dimensional random vector. $\xi_i^\nu$ is an i.i.d Gaussian random
number of zero mean and unit variance. The inequality version of the
problem is referred to as the perceptron and has already been well
investigated~\cite{gardner1989optimal,franz2015universal,franz2016simplest,franz2017}.
It is known that the perceptron exhibits a sharp phase transition from
the satisfiable (SAT) phase, where all constraints are satisfied, to
the unsatisfiable (UNSAT) phase, where some constraints are
violated~\cite{gardner1989optimal}. Later, we show that our model also
exhibits a similar SAT-UNSAT transition. To solve the problem,
we consider the quadratic cost function:
\begin{align}
 V_N = \frac{1}{2N}\sum_{\nu=1}^M (\bx\cdot\bxi^\nu)^2 + \frac{\mu}{2}(N-\bx\cdot\bx),\label{102404_16Aug22}
\end{align}
where $\mu$ denotes the Lagrange multiplier to impose the spherical
constraint $\bx\cdot\bx=N$. The cost function Eq.~(\ref{102404_16Aug22})
can be considered as a special case of the linealized
perceptron~\cite{hertz1989dynamics,watkin1993} with the spherical
constraint. Eq.~(\ref{102404_16Aug22}) is also very similar to that of
the perceptron of $\sigma=0$, where the cost function of the model is
convex~\cite{franz2017}. Later, we show that the current model indeed
exhibits the same scaling as that of the perceptron of $\sigma=0$. When
the conditions Eqs.~(\ref{100328_16Aug22}) are satisfied, one obtains
$V_N=0$ and vice versa.  After some manipulations,
Eq.~(\ref{102404_16Aug22}) is rewritten as
\begin{align}
 V_N 
 =\frac{1}{2}\bx\cdot W \cdot \bx + \frac{\mu}{2}(N-\bx\cdot\bx),\label{102614_16Aug22}
\end{align}
where $W$ is a $N\times N$ symmetric matrix whose $ij$
component is given by
\begin{align}
W_{ij} = \frac{1}{N}\sum_{\nu=1}^{M} \xi_i^\nu \xi_j^\nu.
\end{align}
To investigate the model, we diagonalize the
matrix $W$ and expand the potential by the normal modes:
\begin{align}
 V_N = \sum_{i=1}^N \frac{\lambda_i-\mu}{2}u_i^2 +\frac{\mu}{2}N,\label{110746_16Aug22}
\end{align}
where $\lambda_i$ denotes the $i$-th eigenvalue of $W$.
We will order $\lambda_i$ such that
\begin{align}
 \lambda_1 < \lambda_2 < \cdots < \lambda_N.
\end{align}
Since $W$ is a Wishart matrix, in the thermodynamic limit $N\to\infty$,
its distribution is given by the Marcenko-Pastur
law~\cite{franz2015universal,livan2018introduction}:
\begin{align}
 &\rho(\lambda) = \theta(1-\alpha)(1-\alpha)\delta(\lambda)
 + g(\lambda)\new
& g(\lambda) =
\begin{cases}
 \frac{1}{2\pi}\frac{\sqrt{(\lambda-\lambda_{-})(\lambda_+-\lambda)}}{\lambda} &
 \lambda\in [\lambda_{-},\lambda_+],\\
 0 & {\rm otherwise}
\end{cases},\label{113511_16Aug22}
\end{align}
where $\theta(x)$ denotes the Heaviside step function, ${\alpha=M/N}$
denotes the number of the constraints per degree of freedom, and
\begin{align}
\lambda_{\pm} = (\sqrt{\alpha}\pm 1)^2.\label{120549_7Sep22}
\end{align}
It is easy to show that the ground state energy of
Eq.~(\ref{110746_16Aug22}) is given by
\begin{align}
 \frac{V_{\rm GS}}{N} = \frac{\lambda_{\rm min}}{2} = 
 \begin{cases}
  0 & \alpha <1\\
  \frac{(\sqrt{\alpha}-1)^2}{2} & \alpha >1
 \end{cases},
\end{align}
where $\lambda_{\rm min}$ denotes the minimal eigenvalue of $W$. When
$\alpha<1$, $V_{\rm GS}=0$, implying that all constraints
Eq.~(\ref{100328_16Aug22}) are satisfied. On the contrary, when
$\alpha>1$, $V_{\rm GS}>0$, implying that some constraints are
unsatisfied. Therefore, the model exhibits the SAT-UNSAT transition at
$\alpha_c=1$~\cite{hertz1989phase}. At the transition point, the system
is isostatic: the number of degrees of freedom is the same as that of
the constraints $N=M$~\cite{franz2017}. The isostaticity has been
previously reported for the perceptron for $\sigma\leq
0$~\cite{franz2016simplest,franz2017}.

Now we characterize the criticality around $\alpha_c$. For this purpose,
we first consider the model in equilibrium at temperature $T$ and take
the limit $T\to 0$ at the end of the calculation. From the equipartition
theorem~\cite{greiner2012thermodynamics}, we get
\begin{align}
\ave{u_i^2} = \frac{k_BT}{\lambda_i-\mu},\label{145243_16Aug22}
\end{align}
where $k_B$ denotes the Boltzmann constant. Hereafter, we set $k_B=1$ to
simplify the notation.
Since $\ave{u_i^2}\geq 0$, $\mu$ should satisfy
\begin{align}
\mu\leq \lambda_{\rm min}.
\end{align}
Since an orthogonal transformation preserves the inner product,
the spherical constraint $\bx\cdot\bx=N$ is written as 
\begin{align}
1=\frac{\bx\cdot\bx}{N}= \frac{\bu\cdot\bu}{N}
= \int_{-\infty}^{\infty} d\lambda \rho_N(\lambda)\frac{T}{\lambda - \mu},\label{124932_16Aug22}
\end{align}
where we have defined the distribution of $\lambda_i$:
\begin{align}
 \rho_N(\lambda) = \frac{1}{N}\sum_{i=1}^N\delta(\lambda-\lambda_i).
\end{align}
In the limit $N\to\infty$, $\rho_N(\lambda)$ converges to
Eq.~(\ref{113511_16Aug22}). Therefore, we get
\begin{align}
 1 = -(1-\alpha)\theta(1-\alpha)\frac{T}{\mu}
 + \int_{\lambda_-}^{\lambda_+}d\lambda g(\lambda)\frac{T}{\lambda-\mu}.
\end{align}
We first investigate the behavior of the model in the SAT phase
($\alpha<1$). For $T\ll 1$, the dominant contribution comes from
the first term, thus we get
\begin{align}
 1 = -\frac{(1-\alpha)T}{\mu}\to \mu = -T(1-\alpha),\label{160313_16Aug22}
\end{align}
and 
\begin{align}
\ave{u_i^2} =
 \begin{cases}
 (1-\alpha)^{-1} & i=1,\dots, (1-\alpha)N\\
 0  & {\rm otherwise}
 \end{cases}.\label{160328_16Aug22}
\end{align}
As we approach the transition point $\alpha=1$, the relaxation time
$\tau$ would diverge.  $\tau$ is controlled by the slowest mode, which
has the smallest curvature $\kappa = \min_{i}\partial_{u_i}^2 V_N =
\partial_{u_1}^2 V_N$ along that direction.  Assuming the exponential
decay $\dot{u}_1(t) \propto -\kappa u_1(t)$, $\tau$ is estimated as
\begin{align}
\tau \sim \ave{\pdiff{^2V_N}{u_1^2}}^{-1} = (-\mu)^{-1} = T^{-1}(1-\alpha)^{-1}.\label{173823_16Aug22}
\end{align}
Therefore, the relaxation time diverges as $\tau\sim
(1-\alpha)^{-\beta}$ with the critical exponent $\beta=1$.  The result
seems to be consistent with the previous research for the linealized
perceptron with the Langevin dynamics~\cite{hertz1989dynamics}. For the
perceptron, to the best of our knowledge, the relaxation time in the SAT
phase has not been calculated yet, due to the complexity of the
dynamical equation~\cite{hwang2020,folena2022introduction}.
\begin{figure}[t]
\begin{center}
\includegraphics[width=9cm]{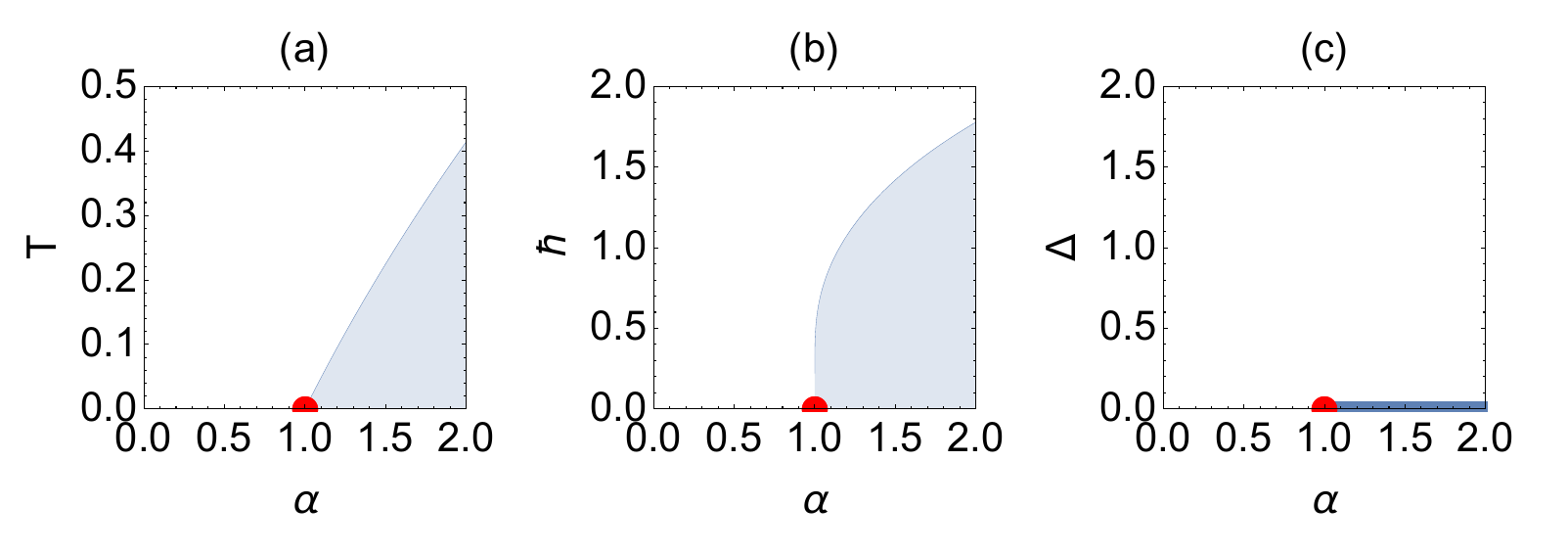} \caption{Phase diagrams for (a)
 classical model, (b) quantum model, and (c) model with random
 field. The filled region represents the spin-glass phase, and the red
 point represents the SAT-UNSAT transition point. For the model with the
 random field, the spin-glass phase does not appear for $\Delta>0$.}
 \label{171415_17Aug22}
\end{center}
\end{figure}

For $\alpha> 1$, $\mu$ is to be determined by 
\begin{align}
 1= F(\mu),\label{123353_16Aug22}
\end{align}
where
\begin{align}
 F(\mu) = T\int_{\lambda_-}^{\lambda_+}d\lambda \frac{g(\lambda)}{\lambda-\mu}.
\end{align}
One can show that $F(\mu)$ takes its maximum value at $\mu=\lambda_-$:
\begin{align}
 F(\lambda_-) = \frac{T}{T_c},\label{145330_1Sep22}
\end{align}
where
\begin{align}
 T_c=\left[\int_{\lambda_-}^{\lambda_+}d\lambda \frac{g(\lambda)}{\lambda-\lambda_{-}}\right]^{-1}
 =\sqrt{\alpha}-1,
\end{align}
see Fig.~\ref{171415_17Aug22}~(a). Below $T_c$, however,
Eq.~(\ref{123353_16Aug22}) has no solution, which is the signature of
the condensation to the lowest eigenmode
$\lambda_-$~\cite{gunton1968condensation,dalfovo1999,barbier2022generalised}.
In the case of the $p=2$-spin spherical mode, the condensation
transition occurs as a consequence of the underlying spin-glass
transition~\cite{barbier2022generalised}. Here we assume that the same
is true for our model and identify $T_c$ with the spin-glass transition
point. For $T<T_c$, the first and the other terms in
Eq.~(\ref{124932_16Aug22}) should be treated separately, as in the case
of the Bose-Einstein condensation~\cite{greiner2012thermodynamics}:
\begin{align}
 1 = \frac{\ave{u_1}^2}{N} + \frac{1}{N}\sum_{i=2}^N \ave{u_i^2}
 = \frac{\ave{u_1}^2}{N} + F(\lambda_-).
\end{align}
From the above equation and Eq.~(\ref{145330_1Sep22}), we get
\begin{align}
\frac{\ave{u_1^2}}{N} = 1- \frac{T}{T_c}.\label{154554_1Sep22}
\end{align}
This can be identified with the Edward–Anderson order
parameter~\cite{edwards1975theory,barbier2022generalised}. To
characterize the criticality in the limit $T\to 0$,
following~\cite{franz2017}, we define the susceptibility against the
thermal fluctuation:
\begin{align}
\chi = \lim_{T\to 0}\frac{1}{T}\left[1-\frac{\ave{u_1}^2}{N}\right] = \frac{1}{\sqrt{\alpha}-1}.\label{153546_16Aug22}
\end{align}
The susceptibility diverges as $\chi\propto (\alpha-1)^{-1}$ on
approaching the SAT-UNSAT transition point.  The same critical exponent
has been previously reported for the perceptron for
$\sigma=0$~\cite{franz2017}.

Finally, we comment on the marginal stability.  As in the SAT phase, the
$i$-th eigenvalue $\lambda_i$ of the Hessian of the interaction
potential Eq.~(\ref{110746_16Aug22}) is calculated as
\begin{align}
\tl_i = \ave{\pdiff{V_N}{u_i^2}} = \lambda_i-\mu.\label{120209_7Sep22}
\end{align}
Using $\ave{u_1^2}= 1/(\lambda_{-}-\mu)$ and Eq.~(\ref{154554_1Sep22}),
we get in the spin-glass phase
\begin{align}
\mu = \lambda_{-}- \frac{1}{N}\frac{T_c}{T_c-T}\to \lambda_{-}.
\end{align}
Therefore, the system is marginally stable ${\tl_{\rm min}=
\lambda_{-}-\lambda_{-}=0}$ in the thermodynamic limit
$N\to\infty$~\cite{muller2015marginal}.  In the terminology of the
effective medium theory (EMT), the term $\mu$ that shifts all
eigenvalues, as in Eq.~(\ref{120209_7Sep22}), is referred to as the {\it
pre-stress}~\cite{degiuli2014effects}. Using Eq.~(\ref{120549_7Sep22})
and the marginal stability $\tl_{\rm min}=\lambda_{-}-\mu=0$, one obtain
a well-known square root scaling $\alpha-1 \propto
\sqrt{\mu}$~\cite{degiuli2014effects}. Note that in the case of the EMT,
the marginal stability is an assumption, but in the case of the current
model, the value of $\mu$ is fine-tuned automatically to achieve the
marginal stability as a consequence of the spin-glass transition for
$T<T_c$. The distribution of $\tl_i$ is calculated as
\begin{align}
 \rho(\tl) = \rho(\lambda=\tl+\mu)\diff{\lambda}{\tl}=\rho(\lambda+\lambda_-),
\end{align}
where $\rho(\lambda)$ is given by Eq.~(\ref{113511_16Aug22}).  To
compare with previous researches~\cite{ohern2003,charbonneau2016}, we
define the vibrational density of states: the distribution of the
eigenfrequency $\omega_i=\sqrt{\tl_i}$,
\begin{align}
 D(\omega) &= \rho(\tl=\omega^2)\diff{\tl}{\omega}\new
 &= \frac{\omega^2\sqrt{\lambda_+-\lambda_--\omega^2}}{\pi(\omega^2+\lambda_-)}.
\end{align}
One can deduce the scaling behavior of $D(\omega)$ at $T=0$ near the
transition point $\alpha\approx 1$ as follows:
\begin{align}
 D(\omega)\sim
 \begin{cases}
  {\rm const} & \omega \gg \omega_*\\
  (\omega/\omega_*)^2 & \omega \ll \omega_*
 \end{cases},
\end{align}
where $\omega_*= (\alpha-1)/2$. The same scaling has been previously
derived for the perceptron for $\sigma\leq 0$~\cite{franz2015universal},
the EMT for a disordered lattice~\cite{degiuli2014effects}, and
numerical simulation of sphere packing near the jamming transition
point~\cite{ohern2003,charbonneau2016,hikeda2022}.

\section{Quantum model}
\label{145228_17Aug22}
Now, we consider the quantum version of the model:
\begin{align}
 H = \sum_{i=1}^N\frac{p_i^2}{2} + \sum_{i=1}^N\left[\frac{\lambda_i-\mu}{2}u_i^2\right],
\end{align}
where we omit the constant term $\mu N/2$ to simplify the notation.
We require the standard canonical commutation
relation~\cite{vojta1996,cugliandolo2001}:
\begin{align}
 [u_i,u_j] = 0,\ [p_i,p_j] = 0,\ [u_i,p_j] = \hbar\delta_{ij},
\end{align}
where $\hbar$ is the plank constant. Here we use $\hbar$ as a control
parameter to control the strength of the quantum fluctuation. Following
the standard operation of quantum statistical
mechanics~\cite{greiner2012thermodynamics}, one can calculate the
partition function for the $i$-th harmonic oscillator as
\begin{align}
 Z_i &= \tr e^{-\beta H} = \sum_{n=0}^\infty e^{-\beta\hbar \sqrt{\lambda_i-\mu} (n+1/2)}\new 
 &= \left[2\sinh\left(\frac{\beta\hbar\sqrt{\lambda_i-\mu}}{2}\right)\right]^{-1},
\end{align}
where $\beta=1/T$ denotes the inverse temperature.  Then, the second
moment is
\begin{align}
\ave{u_i^2} = \frac{2}{\beta}\diff{\log Z_i}{\mu}
= \left[\frac{2\sqrt{\lambda_i-\mu}}{\hbar}\tanh
\left(\frac{\hbar\sqrt{\lambda_i-\mu}}{2T}\right)
 \right]^{-1}.\label{145226_16Aug22}
\end{align}
In the high temperature limit, we get
\begin{align}
\frac{2\sqrt{\lambda_i-\mu}}{\hbar}\tanh
\left(\frac{\hbar\sqrt{\lambda_i-\mu}}{2T}\right)\sim \frac{\lambda_i-\mu}{T}.
\end{align}
Substituting it back into Eq.~(\ref{145226_16Aug22}), we recover the
classical result Eq.~(\ref{145243_16Aug22}).  Instead, here we first
take the $T\to 0$ limit and then observe the asymptotic behavior for
${\hbar\ll 1}$.  At $T=0$, Eq.~(\ref{145226_16Aug22}) reduces to
\begin{align}
\ave{u_i^2} = \frac{\hbar}{2\sqrt{\lambda_i-\mu}}.
\end{align}
As before, $\mu$ is determined by the spherical constraint:
\begin{align}
 1 = \frac{1}{N}\sum_{i=1}^N\ave{u_i}^2 = 
\frac{1}{N}\sum_{i=1}^N\frac{\hbar}{2\sqrt{\lambda_i-\mu}}.\label{103921_17Aug22}
\end{align}
Repeating the same analysis of that of the classical model, one can see
that for $\alpha<1$ and $\hbar\ll 1$, Eq.~(\ref{103921_17Aug22}) reduces to
\begin{align}
 1 = \frac{1}{N}\sum_{i=1}^{(1-\alpha)N}\ave{u_i^2}
 = (1-\alpha)\frac{\hbar}{2\sqrt{-\mu}},
\end{align}
which leads to
\begin{align}
&\mu = -\frac{\hbar^2(1-\alpha)^2}{4},\new 
&\ave{u_i^2} =
 \begin{cases}
  (1-\alpha)^{-1} & i=1,\dots, (\alpha-1)N\\
  0 & {\rm otherwise}
 \end{cases}.
\end{align}
We find the same exponent for $\ave{u_1^2}$ and the different exponent
for $\mu$ from those of the classical model, see
Eqs.~(\ref{160313_16Aug22}) and (\ref{160328_16Aug22}). The similar
result has been previously obtained for the perceptron for
$\sigma=0$~\cite{artiaco2021}, where the authors mentioned the
differences in the critical exponent between the classical and quantum
models. For $\alpha>1$, the condensation (spin-glass) transition occurs
at a finite $\hbar=\hbar_c$.  As before, this transition point is
calculated as
\begin{align}
&1 = \frac{\hbar}{2}\int_{\lambda_-}^{\lambda_+} d\lambda \frac{g(\lambda)}{\sqrt{\lambda-\lambda_-}}\new 
&\to \hbar_c = \frac{2}{\int d\lambda g(\lambda)(\lambda-\lambda_-)^{-1/2}},
\end{align}
see Fig.\ref{171415_17Aug22}~(b). In the limit $\alpha\to 1$, $\hbar_c$
vanishes as
\begin{align}
 \hbar_c \propto -\frac{1}{\log(\alpha-1)}.
\end{align}
For $\hbar<\hbar_c$, we define the order parameter 
\begin{align}
\frac{\ave{u_1^2}}{N} = 1-\frac{\hbar}{2}\sum_{i=2}^N \frac{1}{\sqrt{\lambda_i-\lambda_{-}}},
\end{align}
and susceptibility w.r.t the quantum fluctuation:
\begin{align}
 \chi = \lim_{\hbar\to 0}\frac{1}{\hbar}\left(1-\frac{\ave{u_1^2}}{N}\right)
 = \frac{1}{2}\int_{\lambda_-}^{\lambda_+} d\lambda \frac{g(\lambda)}{\sqrt{\lambda-\lambda_-}}.
\end{align}
In the limit $\alpha\to 1$, $\chi$ diverges logarithmically 
\begin{align}
 \chi \sim -\log(\alpha-1),
\end{align}
instead of the power-law found in the classical model
Eq.~(\ref{153546_16Aug22}). To the best of our knowledge, this is the
first result that reveals qualitative differences 
between the thermal and quantum fluctuations near the continuous SAT-UNSAT
transition point.

Finally, we would briefly comment on the marginal
stability~\cite{muller2015marginal}.  As in the case of the classical
model, the minimal eigenvalue of the Hessian is calculated as
\begin{align}
 \tl_1 = \ave{\pdiff{V_N}{u_1^2}} = \lambda_{-}-\mu.
\end{align}
Repeating a similar analysis as in the previous section, one can show
that $\tl_1 >0$ for $\hbar>\hbar_c$, and $\tl_1=0$ for $\hbar\leq
\hbar_c$, implying that the density of states is gapped for
$\hbar>\hbar_c$ even at $T=0$. In particular, at $\alpha=\alpha_c=1$,
the gap vanishes only in the limit $\hbar\to 0$, see
Fig.~\ref{171415_17Aug22}~(b).  This gap was not reported in a previous
calculation for the perceptron based on the Schehr–Giamarchi–Le Doussal
Expansion, which is the expansion by $\hbar$ with fixed
$\hbar/T$~\cite{franz2019impact}. Further studies of the quantum version
of the perceptron with more accurate approximations would be beneficial
to clarify the origin of this discrepancy.

\section{Effects of random field}
\label{145448_17Aug22}

Finally, we consider the model with the random field:
\begin{align}
 V_N = \sum_{i=1}^N \frac{\lambda_i-\mu}{2}u_i^2 + \sum_{i=1}^N h_i u_i,
\end{align}
where $h_i$ is an i.i.d random variable of zero mean and variance $\Delta$.
In equilibrium at temperature $T$, we get
\begin{align}
 \overline{\ave{u_i^2}} = \frac{T}{\lambda_i-\mu} + \frac{\Delta}{(\lambda_i-\mu)^2},
\end{align}
where the overline denotes the average for $h_i$, and
$\Delta=\overline{h_i^2}$. Hereafter we consider the model at $T=0$, and
observe the asymptotic behavior for $\Delta\ll 1$. As before, $\mu$ is
determined by the spherical constraint:
\begin{align}
 1 = \frac{1}{N}\sum_{i=1}^N \frac{\Delta}{(\lambda_i-\mu)^2}.\label{174319_16Aug22}
\end{align}
Repeating the same analysis of that of the classical model,
in the limit $\Delta\ll 1$, we get for $\alpha>1$
\begin{align}
\mu &= -(1-\alpha)^{1/2}\Delta^{1/2},\new 
\ave{u_i^2} &=
 \begin{cases}
  (1-\alpha)^{-1} & i=1,\dots, (\alpha-1)N\\
  0 & {\rm otherwise}
 \end{cases}.
\end{align}
We find the same exponent for $\ave{u_1^2}$ and the different exponent
for $\mu$ from both the classical and quantum models.



The condensation (spin-glass) transition point for $\alpha>1$ is
estimated as
\begin{align}
 \Delta_c = \frac{1}{\int d\lambda g(\lambda)(\lambda-\lambda_-)^{-2}} = 0,
\end{align}
meaning that the transition does not occur at finite $\Delta$, see
Fig.~\ref{171415_17Aug22}~(c). In a previous work, we investigated a
similar equation as Eq.~(\ref{174319_16Aug22}) and found that the
transition at finite $\Delta$ occurs only for $\rho(\lambda)\sim
(\lambda-\lambda_-)^n$ with $n>1$~\cite{ikeda2022bose}, which is not
satisfied by the eigenvalue distribution of the current model
Eq.~(\ref{113511_16Aug22}).

\section{Summary and discussions}
\label{145458_17Aug22} In this work, we investigated the convex
continuous SAT-UNSAT transition of a special case of the linearized
perceptron. Since the interaction potential has a quadratic form, the
model can be easily analyzed by expanding the potential by the normal
modes. We successfully characterized the criticality near the SAT-UNSAT
transition point. The simplicity of the model allows us to investigate
the quantum effects, which have not been fully investigated before. We
found different critical behaviors from those of the classical model. In
particular, the susceptibility of the order parameter diverges
logairthmically when approaching the transition point from the UNSAT
side. This is qualitatively different behavior from that of the
classical model, where the power-law divergence is observed.  Finally,
we investigated the model with the random field. We found different
critical exponents from both the classical and quantum models.


There are still several important points that deserve further
investigation. Here we give a tentative list:
\begin{itemize}
 \item The dynamics of the linearized perceptron has been already well
       investigated~\cite{hertz1989dynamics,cun1991,watkin1993}.
       Furthermore, the simplicity of the model may allow us to derive a
       full dynamical solution as done for the $p$-spin spherical
       models~\cite{cugliandolo1993analytical,cugliandolo1995full}. It
       would be interesting to revisit these results and compare them
       with recent numerical simulations of particle systems near the
       jamming (SAT-UNSAT) transition
       point~\cite{ikeda2020universal,nishikawa2021relaxation,nishikawa2022relaxation}.

\item We found that $D(\omega)$ has a finite gap for $\hbar>0$, which has
       not been reported before~\cite{franz2019impact}. It is
      interesting to investigate how this gap affects the low
       temperature behavior, in particular, 
      the temperature dependence of the specific heat.

\item In this work, we considered the quadratic cost function
Eq.~(\ref{102404_16Aug22}). A
natural generalization is to consider the $p$-body interaction as in the
case of the $p$-spin spherical
model~\cite{crisanti1992sphericalp,castellani2005spin}:
      \begin{align}
       V_N = \frac{1}{N^{p}}\sum_{\nu=1}^M\left(\bx\cdot\bxi^{\nu}\right)^p + \frac{\mu}{2}\left(N-\bx\cdot\bx\right).
      \end{align}
By analogy from the $p$-spin spherical mode, we expect that the model
exhibits the one-step (or higher) replica symmetric breaking (1RSB) for
$p>2$~\cite{castellani2005spin}.  Thus, the phase diagram of the model
is more similar to that of the structural
glasses~\cite{cavagna2009supercooled,charbonneau2014fractal}.  It would
be interesting to investigate how the 1RSB transition affects the
SAT-UNSAT transition.
\end{itemize}

\acknowledgments
This project has received
JSPS KAKENHI Grant Numbers 21K20355.

\bibliography{reference}

\end{document}